\documentclass[superscriptaddress,
			   nofootinbib,
			   aps,pra,
			   twocolumn,10pt]{revtex4-1}

\usepackage[T1]{fontenc} 
\usepackage[english]{babel} 
\usepackage[utf8]{inputenc}

\usepackage{graphicx}

\usepackage[table,usenames,dvipsnames]{xcolor}

\usepackage{hyperref} 
\PassOptionsToPackage{linktocpage}{hyperref} 
\hypersetup{
  colorlinks	= true, 
  urlcolor		= violet, 
  linkcolor		= NavyBlue, 
  citecolor		= Green, 
  hypertexnames	= false,
}

\usepackage{amssymb}
\usepackage{amsmath,mathtools}
\usepackage{amsthm}
\usepackage{dsfont}
 
\newtheorem{theorem}{Theorem}
\newtheorem{corollary}[theorem]{Corollary}
\newtheorem{definition}[theorem]{Definition}

\newcommand{\e}{\ensuremath\mathrm{e}}

\DeclareMathOperator{\landauO}{\mathrm{O}}
\DeclareMathOperator{\landauOmega}{\Omega}
\DeclareMathOperator{\Tr}{Tr}

\DeclareMathOperator{\supp}{supp}
\DeclareMathOperator{\dist}{dist}

\DeclareMathOperator{\cov}{cov}
\newcommand{\rmd}{\mathrm{d}}
\renewcommand{\L}{\operatorname{L}}

\newcommand{\CC}{\mathbb{C}}
\newcommand{\RR}{\mathbb{R}}

\newcommand{\1}{\mathds{1}}

\newcommand{\mc}[1]{\mathcal{#1}}
\renewcommand{\H}{\mc{H}}

\newcommand{\argdot}{{\,\cdot\,}}

\newcommand{\norm}[1]{\left\Vert #1 \right\Vert} 
\newcommand{\sNorm}[1]{\norm{#1}} 

\newcommand{\TrNorm}[1]{\norm{#1}_{1}} 

\newcommand{\gibbs}{g}
\newcommand{\animalc}{\alpha}
\newcommand{\Eset}{\mc{E}}
\newcommand{\Vset}{\mc{V}}
\newcommand{\micro}{\sqcap}

\usepackage{tikz}
\usetikzlibrary{calc,fadings,backgrounds,shapes.geometric,positioning}

\tikzset{lattice/.style = {anchor = center},
	 vertex/.style = {circle,inner sep = 1.2ex},
	 beschriftung/.style = {transform shape = false, text opacity=1},
	 box/.style = {rounded corners =.5ex},
	 edges/.style = {very thick, line cap = round, blue!30},
	 }%
\definecolor{niceblue}{rgb}{0.33,0.5,0.8}%
\colorlet{BboxCol}{gray!10}%
\colorlet{SboxCol}{gray!25}%
\def \nx{14}%
\def \ny{9}%
\def \Vsize{1ex}%
\newcommand{\drawlattice}[2]{%
  \begin{scope}%
  \def \nx{#1}%
  \def \ny{#2}%
  \clip ($(\nx,\ny) + (-.5,-.5)$) rectangle (.5,.5);
  \foreach \x in {0,1,...,\nx}
    \foreach \y in {0,1,...,\ny}
      \node (v\x_\y) [vertex] at (\x,\y){}; 
  \foreach \x in {1,2,...,\nx}{
    \pgfmathsetcount{\xm}{\x-1};
    \foreach \y in {1,2,...,\ny}{
      \pgfmathsetcount{\ym}{\y-1};
      \draw[edges] (v\x_\y) -- (v\the\xm_\y)
      (v\x_\y) -- (v\x_\the\ym);
    }%
  }%
  \foreach \x in {0,1,...,\nx}
    \foreach \y in {0,1,...,\ny}{
      \draw[gray] (\x,\y) circle (\Vsize); 
      \shade[ball color=gray] (\x,\y) circle (\Vsize);
  }%
\end{scope}
}%
\newcount\xm%
\newcount\ym%
\newcount\xx%
\newcount\yy%

\newcommand{\ug}{Institute of Theoretical Physics and Astrophysics, 
		  University of Gda\'nsk, 
		  Poland}

\newcommand{\hhu}{
	Institute for Theoretical Physics,
	Heinrich Heine University D{\"u}sseldorf, 
	Germany
}

\newcommand{\icfo}{ICFO -- Institut de Ciencies Fotoniques, The Barcelona Institute of Science and Technology, 
			Castelldefels, Spain}

\newcommand{\mpq}{Max-Planck-Institut f\"ur Quantenoptik, 
			Garching, Germany}

\makeatletter 
 \hypersetup{pdftitle = {Properties of thermal states: locality of temperature, decay of correlations, and more},
	     pdfauthor = {Martin Kliesch, Arnau Riera},
	     pdfsubject = {Quantum physics},
	     pdfkeywords = {Gibbs state, quantum, 
	     				thermal states, 
	     				clustering, 
	     				exponential, decay, 
	     				energy distribution, 
	     				equivalence of ensembles,
	     				locality, inverse, temperature,
	     				Lieb-Robinson, upper, bound, 
	     				book chapter, Thermodynamics in the Quantum Regime,
	     				}
	    }
\makeatother

\begin{document}

\title{Properties of thermal states:
	   \texorpdfstring{\\}{} 
	   locality of temperature, decay of correlations, and more}

\author{Martin Kliesch}
\email{science@mkliesch.eu}
\affiliation{\ug}
\affiliation{\hhu}

\author{Arnau Riera}
\email{arnauriera@gmail.com} 
\affiliation{\icfo}
\affiliation{\mpq}

\begin{abstract}
We review several properties of thermal states of spin Hamiltonians with short range interactions.
In particular, we focus on those aspects in which the application of tools coming from quantum information theory
has been specially successful in the recent years.
This comprises the study of the correlations at finite and zero temperature, the stability against distant and/or weak perturbations,
the locality of temperature and their classical simulatability.
For the case of states with a finite correlation length,
we overview the results on their energy distribution and the equivalence of the canonical and microcanonical ensemble.
\end{abstract}

\maketitle

\section{Introduction} \label{sec:intro}
Thermal or Gibbs states are arguably the most relevant class of states in nature.
They appear to have many very different unique properties. 
On one hand, they provide an efficient description of the equilibrium state
for most systems (see 
\cite[Chapter~17]{BinCorGog19}
and Ref.~\cite{GogEis15} for a detailed review).
Jaynes gave a justification of this observation by means of the principle of maximum entropy \cite{Jaynes1957, Jaynes1957a}.
For a given energy, the thermal state is the one that maximizes the von Neumann entropy.
This is in fact the reason why they minimize the free energy potential.

At the same time, in the context of extracting work from closed systems by means of unitary transformations, thermal states appear as the only ones that are
completely passive states \cite{AlickiFannes13}. That is, for any other passive state \cite{Pusz1978}, i.e. a state whose (average) energy cannot be lowered by applying a unitary transformation, 
it is always possible to form an active state by joining sufficiently many copies.
Note that this property is crucial for establishing both a resource theory of thermodynamics and a consistent zero-th law
(see the book chapters 
\cite[Chapters~25]{BinCorGog19} and
\cite{NatWinLew18}
).

Last but not least, if one accepts that the equilibrium population of a configuration should only be a function of its energy, the canonical ensemble becomes the single probability distribution that allows for a free choice of the energy origin, or in other words, that is invariant under energy shifts. 
More explicitly, any other relation between populations and energy would imply that observers with different energy origins would observe different equilibrium states. 
The proof of this statement is a good exercise. 

In this chapter, we study of the properties of thermal states
for a class of Hamiltonians highly relevant in condensed matter physics: 
the spin lattice Hamiltonians with short range interactions.
We particularly focus on results where the application of tools coming from quantum information theory
has been particularly successful in the recent years.

A first issue where quantum information tools have been helpful
is in the rigorous proof of long-standing conjectures concerning the correlations of many-body systems. 
The most paradigmatic example is the proof that unique ground states of gapped
Hamiltonians have an exponential decay of the correlations \cite{HasKom06,NacSim06}.
In Section~\ref{sec:correlations} we overview the main results on 
the scaling of the correlations for both thermal states and absolute zero temperature states.
The understanding of the correlations of many-body systems is particularly relevant since they are a signature of phase transitions and critical phenomena \cite{Cardy1996scaling,Domb2000phase}.
Furthermore, they are also related to the stability of thermal states, that is, to the robustness of thermal states against distant and/or weak Hamiltonian perturbations \cite{KliGogKas14}. 

The stability of thermal states is actually crucial 
for understanding to what extent temperature can be assigned locally in a system \cite{KliGogKas14,Senaida2015}.
For weakly interacting systems temperature is an intensive quantity, in the sense that every subsystem is in a thermal state with the temperature identical to the global one. 
However, when interactions between subsystems are not negligible, this property does no longer hold, as can be the case for quantum systems \cite{HarMah05,Har06}.
Further steps to circumvent the problem of assigning temperature locally to a
subsystem were made in Refs.~\cite{FerGarAci12,GarFerAc2009}, where it was shown that it is sufficient to extend the subsystem 
by a boundary region that, when traced out, disregards the correlations and the boundary effects. 
If the size of such a boundary region is independent of the total system size, temperature can still be said to be \emph{local}.
In Ref.~\cite{KliGogKas14} it is shown that the thickness of the boundary region is determined by the decay of a \emph{generalized covariance}, which captures the response in a local
operator of perturbing a thermal state and ultimately at what length scales temperature can be defined.
In other words, the clustering of the generalized covariance allows
for a local definition of temperature.
This issue is particularly relevant in 
the book chapter \cite{PasSta18}, where the problem of local thermometry is considered. 

Another feature of the thermal states that exhibit a clustering of correlations is that they can be efficiently classically represented by 
so-called tensor network states \cite{MolSchVer14}. 
The simulation of quantum many body systems by classical means, and more specifically by using tensor network states, has been one of the most fruitful topics in the interplay between condensed-matter physics and quantum information.
Tensor networks have been shown to satisfactorily describe locally entangled states in one and two spatial dimensions and even systems at criticality \cite{Sch11}.
Applications 
of thermal states with exponentially decaying correlations are presented in Section~\ref{sec:applications}.

In Section~\ref{sec:energy_distribution}, we focus on the study of the energy distribution of states not necessarily thermal but with a finite correlation length.
In particular, it can be shown that for large system sizes the energy distribution of a state with a finite correlation length has an energy distribution that tends to a Gaussian whose standard deviation scales with the square root of the system size \cite{BraCra15}. 

The last part of the chapter in Section~\ref{sec:EoE} is dedicated to present a rigorous formulation of the equivalence between the canonical and the microcanonical ensembles \cite{MueAdlMas13,BraCra15}.

We finally conclude with a brief summary and a discussion on what might be the most relevant open questions of the field.

\section{Setting}
Throughout the chapter we set $\hbar=1=k_{\mathrm{Boltzmann}}$ and assume all Hilbert space dimensions to be finite. 
For a positive integer $n$ we set $[n]\coloneqq \{1,2,\dots, n\}$. 
We use the Landau symbols $\landauO$ and $\landauOmega$ to denote upper and lower bounds, respectively, and $\tilde \landauO$ and $\tilde \landauOmega$ if those bounds hold only up to logarithmic factors. 
The \emph{spectral norm} and \emph{trace norm} on operators are denoted by $\sNorm{\argdot}$ and $\TrNorm{\argdot}$. 
They are given by the largest and the sum of the operator's singular values, respectively. 

We consider quantum spin-lattice systems: 
the Hilbert space is given by 
$\H = \bigotimes_{x \in \Vset} \H_x$, where $\Vset$ is a finite set, called the \emph{vertex set}. 
$\H_x$ are the \emph{local} Hilbert spaces. 
Usually, they have all the same dimension, i.e., $\H_x \cong \CC^d$ and $d$ is called the \emph{local Hilbert space dimension}.  

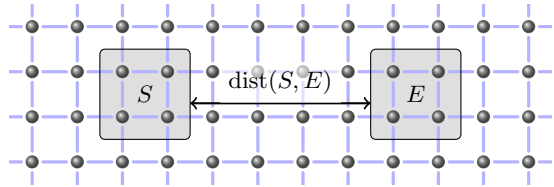
\begin{figure}
	\centering
	\begin{tikzpicture}[lattice,scale = .6,transform shape]
	\def \Sx{3}%
	\def \Sy{2}%
	\def \Sxsize{2}%
	\def \Sysize{2}%
	\def \Bx{9}%
	\def \By{2}%
	%
	\fill [box,SboxCol] (\Sx,\Sy) ++ (-.5,-.5) rectangle ++(\Sxsize,\Sysize);%
	\fill [box,SboxCol] (\Bx,\By) ++ (-.5,-.5) rectangle ++(\Sxsize,\Sysize);%
	%
	\drawlattice{13}{5}%
	%
	\draw [box, draw=black] (\Sx,\Sy) ++ (-.5,-.5) rectangle ++(\Sxsize,\Sysize);
	\path (\Sx,\Sy) ++ (.5,.5) node [beschriftung] (s2) {$S$};
	\draw [box, draw=black] (\Bx,\By) ++ (-.5,-.5) rectangle ++(\Sxsize,\Sysize);
	\path (\Bx,\By) ++ (.5,.5) node [beschriftung] (s2) {$E$};
	%
	\draw [<->, thick] (\Sx, \Sy) ++ (1.5,.3) -- ++(4,0)
	  node[midway, above, fill = white, opacity = .6, beschriftung] {$\dist(S,E)$};
	\end{tikzpicture}%
	\caption{
	Two spatially separated regions $S$ and $E$ in a lattice. 
	}
	\label{fig:covSE}
\end{figure}

The \emph{partial trace} over a subsystem $S\subset \Vset$ of an operator $\rho$ is denoted by $\Tr_S[\rho]$. 
Complementary, we set the reduction of $\rho$ to $S$ to be $\rho^S \coloneqq \Tr_{\Vset \setminus S}[\rho]$.   
The \emph{support} $\supp(A) \subseteq \Vset$ of an observable $A$ on $\H$ is the set of vertices where $A$ does not act as the identity operator. 
For two observables $A,B$ we set their \emph{distance} $\dist(A,B)$ to be the graph distance between their supports; see Figure~\ref{fig:covSE} for an illustration with $S= \supp(A)$ and $E=\supp(B)$. 

%
%

Let us consider a graph $(\Vset,\Eset)$, where $\Eset$ is the \emph{edges set} containing two-element subsets of $\Vset$.  
For a subset $S\subset \Vset$ we denote the set of boundary vertices of $S$ by $\partial S \coloneqq \{ v \in S: \dist(v,\Vset \setminus S) = 1 \}$ and, for an observable $A$ we set  
$\partial A \coloneqq \partial \supp(A)$. 

We say that a Hamiltonian $H$ has \emph{interaction graph} $(\Vset,\Eset)$ and \emph{interaction strength} (bounded by) $J$ if it can be written as
\begin{equation}\label{eq:loc_H}
	H = \sum_{e \in \Eset} h_e
\end{equation}
with $h_e$ being supported on $e$ and $\sNorm{h_e} \leq J$ for all $e \in \Eset$. 

To give an example, for the simple one-dimensional Ising model the introduced quantities are 
$\Vset = [n]$, 
$\Eset = \{ \{1,2\}, \{2,3\}, \dots, \{n-1,n\} \}$, and
$h_{\{j,j+1\}} = \1^{\otimes (j-1)} \otimes \sigma^{(z)}\otimes \sigma^{(z)} \otimes \1^{\otimes (n-j-1)}$, where $\1^{\otimes k}\coloneqq \1\otimes \1\otimes \dots \otimes \1$ denotes $k$ tensor copies of the local identity operator $\1$. 
The support of $h_{\{j,j+1\}}$ is $\{j,j+1\}$. 
As an example for a boundary we have $\partial \{ 2,3,4\} = \{ 2,4\}$. 
Moreover, $\dist(h_{\{1,2\}}, h_{\{6,7\}}) = 6-2 = 4$. 

We denote the \emph{thermal state}, or Gibbs state, of a Hamiltonian $H$ at \emph{inverse temperature} $\beta$ by 
\begin{equation}
  \gibbs(\beta) \coloneqq \frac{\e^{-\beta\,H}}{Z(\beta)} \, ,
\end{equation}
with $Z(\beta) \coloneqq \Tr(\e^{-\beta\,H})$ being the \emph{partition function}. 
If we mean the thermal state or partition function of a different Hamiltonian $H'$, we write $\gibbs[H'](\beta)$ or $Z[H'](\beta)$. 

We measure correlations by the \emph{(generalized) covariance} that we define for any two operators $A$ and $A'$, full-rank quantum state $\rho$, and parameter $\tau \in [0,1]$ as
\begin{equation}\label{eq:def_cov}
 \cov_\rho^\tau(A,A') 
 \coloneqq \Tr\left(\rho^\tau A\, \rho^{1-\tau} A'\right) - \Tr(\rho\, A) \Tr(\rho \, A') \, .
\end{equation}
This quantity is closely related to the Duhamel two-point function, 
which is defined as the average $\int_0^1\rmd \tau \Tr[\rho^\tau A\, \rho^{1-\tau} A']$ 
(see, e.g., \cite[Section~2]{DysLieSim76}). 
The usual \emph{covariance}, denoted by $\cov_\rho(A,A')$, is obtained by setting $\rho^\tau = \rho$ and $\rho^{1-\tau}=\1$, see also \cite[Section~IV.A]{KliGogKas14}. 

It is interesting to point out that the clustering of the standard covariance implies the clustering of the Duhamel two-point function \cite[Theorem 2.1]{DysLieSim76}. 
In this case, we mean by clustering that the covariance tends to zero in the limit of the spatial distance between the non-trivial support of the operators $A$ and $A'$ tends to infinity. 
However, given a state $\rho$ and two observables $A$ and $A'$, the generalized covariance as a function of $\tau$ is not monotonic and can have zeros and saddle points. 

Let us finally introduce the \emph{relative entropy} between two 
quantum states $\rho$ and $\sigma$, which is defined as
\begin{equation}
 S(\rho\|\sigma) \coloneqq - \Tr[\rho \log_2(\sigma)]- S(\rho), 
\end{equation}
where $S(\rho) \coloneqq -\Tr[\rho \log_2(\rho)]$ is the von Neumann entropy. The relative entropy provides a notion of non-symmetric distance between states, and can be operationally interpreted as the physical distinguishability from many copies \cite{AudMosVer12}. 
The relative entropy is closely related to the \emph{free energy}
when its second argument $\sigma$ is a thermal state, 
\begin{equation}
  \label{eq:freeenergydifference}
  F(\rho) - F(\gibbs(\beta)) = \frac{1}{\beta} S(\rho\|\gibbs(\beta)) ,
\end{equation}
where $F(\rho,H)\coloneqq \Tr(\rho H)-\beta^{-1}S(\rho)$ is the out of equilibrium \emph{free energy}.
In a similar way, the \emph{mutual information} of a bipartite state
 $\rho$ 
 can also be written in terms of the relative entropy
 \begin{equation}
I(A:B)_\rho= S(\rho\|\rho_A\otimes\rho_B)\, .
 \end{equation}
Recall that the mutual information, defined as $I(A:B)_\rho\coloneqq S(\rho_A)+S(\rho_B)- S(\rho)$, is a measure of the total correlations between the parties $A$ and $B$ \cite{NieChu10}.

\section[Clustering]{Clustering of correlations}\label{sec:correlations}
In quantum lattice models with local Hamiltonians, a relevant class of observables are the local ones, that is, operators that only have non-trivial support on a few connected sites. 
However, local operators cannot witness global properties of the state
such as entanglement, correlations or long range order.
The simplest observables that can reveal this type of features are the two-point correlation functions or, as they have been introduced in the former section, the covariance.
In condensed-matter physics, the scaling of the covariance with the distance between the two ``points'' is an important correlations measure.
In this section, we review the current understanding on the sorts of
scalings that the covariance of thermal states and zero temperature states can exhibit.

\subsection{High temperature states}
Bounding correlations in thermal spin systems has a long tradition in theoretical and mathematical physics; 
see e.g.\ Ref.~\cite{Rue69} classical systems, \cite{Gin65} for quantum gases (translation-invariant Hamiltonians in the continuum), and Refs.~\cite{BraRob97,Gre69,ParYoo95} cubic lattices. 
Here, we review a general result for finite dimensional systems. 

Any graph that is a regular lattice has an associated (cell or lattice animal) \emph{growth constant} $\animalc \in \RR_+$, which captures the local connectivity of a lattice \cite{Kla67,MirSla11}. 
For instance, for lattices in $D$ spatial dimensions, 
$\animalc \leq 2\,D\,\exp(1)$ \cite[Lemma~2]{MirSla11}. 

The following theorem provides a universal upper bound to all critical temperatures that only depends on the growth constant $\animalc$. 

\begin{theorem}[High temperatures {\cite[Theorem~2]{KliGogKas14}}] 
\label{thm:KGKclustering}
For a graph $(V,\Eset)$ with growth constant $\animalc$
define the constant (critical inverse temperature)
\begin{equation}\label{eq:crit_beta_def}
\beta^\ast 
\coloneqq 
\ln\bigl[\bigl(1+\sqrt{1+4/\animalc}\bigr)/2\bigr]/(2\,J) 
\end{equation}
and function $\xi$ (correlation length) by
\begin{equation} \label{eq:correlation_length_def}
\xi(\beta) 
\coloneqq 
\bigl|\ln\bigl[\animalc\, \e^{2\,|\beta|\,J}\bigl(\e^{2\,|\beta|\,J}-1\bigr)\bigr]\bigr|^{-1}  \, .
\end{equation}
Then, a thermal state $\gibbs$ of any Hamiltonian with interaction graph $(\Vset,\Eset)$ and local interaction strength $J$ satisfies the following: 
for every $|\beta|<\beta^\ast$, parameter $\tau \in [0,1]$, and every two operators $A$ and $B$  
\begin{equation} \label{eq:KGKclustering} 
 |\cov^\tau_{\gibbs(\beta)} (A, B)| 
 \leq 
 \frac{4 \, a\, }  {\ln(3)\, (1-\e^{-1/\xi(\beta)})} \, 
 \e^{-\dist(A,B)/\xi(\beta)} 
\end{equation}
whenever
$\dist(A, B)$ exceeds some minimal distance $L_0$ 
(given in \cite[Eq.~(50)]{KliGogKas14})
and with parameter $a \coloneqq \sNorm{A}\, \sNorm{B} \min\{|\partial A |,|\partial B|\}$.
\end{theorem}

This theorem guarantees that below an inverse universal critical temperature $\beta^\ast$ correlations always decay with a correlation length at most $\xi(\beta)$. 
The inverse critical temperature $\beta^\ast$ and correlation length $\xi$ are universal bounds to the corresponding actual physical quantities. 

For the proof of Theorem~\ref{thm:KGKclustering}, four weighted copies of the system $\Vset$ are considered and a cluster expansion analysis from Hastings \cite{Has06} is tightened for this particular case. 
Originally, Hastings used this cluster expansion to show that high temperature thermal states on spin lattices can be approximated by so-called MPOs; see also Section~\ref{sec:class_sim} on classical simulatability below. 

\subsection{Ground states of gapped Hamiltonians}
The following theorem, first proved by Hastings, Koma, Nachtergaele, and Sims \cite{HasKom06,NacSim06}, 
confirmed a long-standing conjecture in condensed-matter physics, that gapped Hamiltonian systems have exponentially clustering correlations in the ground state. 
A \emph{gapped Hamiltonian} is a Hamiltonian whose gap, i.e., the energy difference between the ground state and the first excited state (or eigenspace), is lower bounded by a constant in the thermodynamic limit. 
The proof of the theorem is based on Lieb-Robinson bounds which provide a bound on the speed of propagation of correlation under Hamiltonian evolution. 
There is a constant speed limit that only depends on the maximum number of lattice nearest neighbors and the interaction strength and is called \emph{Lieb-Robinson speed}, see Ref.~\cite{KliGogEis14} for a review on that topic. 

\begin{theorem}[Unique ground states {\cite[Theorem~4.1]{NacSim10}}]
\label{thm:UniqueGS}
  Let $H$ be a local Hamiltonian with interaction graph $(\Vset, \Eset)$, 
  a unique ground state $\psi$, and 
  a spectral gap $\Delta E>0$. 
  Then, for all observables $A$ and $B$ 
\begin{equation}\label{eq:Clustering_gs}
  \left| \cov_\psi (A, B) \right|
  \leq C\, a \, \e^{-\mu \dist(A,B)}\, ,
\end{equation}
where 
$a \coloneqq \sNorm{A}\, \sNorm{B}_\infty \min\{|\partial A |,|\partial B|\}$, 
the constant $C$ depends on $\Delta E$ and the local lattice geometry, 
and the constant $\mu$ on $\Delta E$ and the Lieb-Robinson speed. 
\end{theorem}

This theorem tells us that correlations in ground states of gapped Hamiltonians decay with a correlation length that is bounded in terms of the spectral energy gap above the ground state. 

It is interesting to point out that 
one dimensional systems in a pure state with exponentially decaying correlations obey an area law for the entanglement entropy \cite{Brandao2015AreaLaw}, that is, the entropy of the reduced state of a subregion grows like its boundary and not like its volume \cite{Eisert2010}. 
The same statement remains open in higher dimensions.
Theorem \ref{thm:UniqueGS} together with Ref.~\cite{Brandao2015AreaLaw}
is an alternative way to see that ground-states of gapped Hamiltonians
obey an area law for the entanglement entropy \cite{Hastings2007}.

For the case of non-zero temperature, 
by inserting in Eq.~\eqref{eq:freeenergydifference} the
thermal states of an interacting Hamiltonian $H=H_S\otimes \1+\1 \otimes H_E+H_I$ 
and a non-interacting one $H_S\otimes \1+\1 \otimes H_E$ of a composite system with parties $S$ and $E$,
one obtains
\begin{equation}
I(S:E)_{\gibbs(\beta)}\leq 2 \beta \sNorm{H_I} - S\bigl[\gibbs^S(\beta) \otimes \gibbs^E(\beta) \big\| \gibbs(\beta)\bigr]
\end{equation}
for the mutual information of $S$ and $E$. 
This equation implies an area law for the mutual information when applied on local Hamiltonians since $\sNorm{H_I}$ scales as the boundary between $S$ and $E$ \cite{Wolf2008} and the relative entropy is non-negative.

The area law for the mutual information does not contradict the fact that most energy eigenstates of non-integrable/chaotic systems away from the edges of the spectrum exhibit a volume law for the entanglement entropy 
\cite{Eisert2010,DasSha06,Masanes2009}.
At relatively large temperatures, when these excited states start to have non-negligible populations,
their contribution to the entropy due to exhibiting a volume law is compensated by the entropy of the global state, which also exhibits a volume law for the entanglement.

\subsection{Critical points at zero and finite temperature}
We have seen that in the regime of high temperatures and for gapped Hamiltonians at absolute zero temperature the covariance decays exponentially with the distance. 
In the following, we review the behavior of correlations
for gapless Hamiltonians at zero temperature and for arbitrary local Hamiltonians below the critical temperature \eqref{eq:crit_beta_def}.

\subsubsection{Gappless Hamiltonians at zero temperature}
If the system is gapless and no other assumption is made on the Hamiltonian, then, roughly speaking, any correlation behavior is possible.
For instance, by a proper fine-tuning of the coupling constants of a nearest neighbor interacting spin-1/2 Hamiltonian, it is possible to maximally entangle spins arbitrarily far away \cite{Vitagliano2010}.
It also remarkable that even ground-states of one dimensional translation invariant Hamiltonians with nearest neighbor interactions can exhibit a volume law for the entanglement entropy \cite{Aharonov2009,Gottesman2009}.

However, in practice, the relevant gapless models in condensed-matter physics have ground states satisfying an area-law for the entanglement entropy with only logarithmic corrections. 
These logarithmic corrections are a manifestation of a power-
law decay in the two-body correlations.
It is useful to introduce the notion of critical system and the critical exponents. 
A system at zero temperature is said to be critical when the spectral gap $\Delta$ between the energy ground state (space) and
the first excited state closes in the thermodynamic limit. 
More explicitly, we say the system to be \emph{critical} when the gap scales as 
\begin{equation}
\Delta \propto N^{-z\nu}\, ,
\end{equation}
where $N=|\Vset|$ is the system size
and $z$ and $\nu$ are the critical exponents that control how fast the gap $\Delta$ tends to zero. 
Actually, $\nu$ is critical exponent that controls the divergence of the correlation length in the vicinity of the critical point, 
\begin{equation}\label{eq:corr-length-div}
\xi \propto N^{\nu}\, ,
\end{equation}
and $z$ the one that determines the dynamic scaling (see Ref.~\cite[Chapter~8]{Cardy1996scaling}). 
The previous divergences are a signature of the \emph{scale invariance} that the system experiences at criticality \cite{Cardy1996scaling}. 
For a critical exponent $z=1$ time and space correlations scale identically. 
This implies a further symmetry enhancement
and the system becomes \emph{conformal invariant}.
The group of conformal transformations includes, in addition to scale transformations, translations and
rotations. 
Such group is particularly powerful in $1+1$ dimensions when we address the problem of the locality of temperature. 
The conformal symmetry completely determines 
the difference between a local expectation value
of the infinite Hamiltonian and the truncated one. 

\subsubsection{Arbitrary systems at finite temperature}
In a similar way to the ground state case, it is an open question
what kind of scalings for the correlation functions of thermal states at finite can exhibit in general.
The most common picture is that there exists at least a renormalization group fixed point in the space of couplings of the Hamiltonians~\cite[Chapter~3]{Cardy1996scaling}. 
Away from the fixed point, the correlations decay exponentially.
In contrast, in its vicinity, the renormalization group transformations can be linearized and some scaling laws for both the free energy and the correlation functions are obtained.
In particular, the correlation length is shown to diverge as in Eq.~\eqref{eq:corr-length-div}.

\subsection[Stability]{Correlations and stability}
Let us again assume that a quantum system is in a thermal state $\gibbs(\beta)$ w.r.t.\ an arbitrary Hamiltonian $H$. 
If we assume that a subsystem $S$ is little correlated with a disjoint subsystem $E$, say due spatial separation as sketched in Figure~\ref{fig:covSE}, then it is expected that a ``perturbation'' of the Hamiltonian on $E$ has only little effect on the state on $S$. 
This expectation can be rigorously confirmed and is implied by the following statement, which holds in more generality (no spin-lattice setup required). 

\begin{theorem}[Perturbation formula {\cite[Theorem~1]{KliGogKas14}}]
\label{thm:perturbation_formula}
Let $H_0$ and $H$ be Hamiltonians acting on the same Hilbert space. 
For $s \in [0,1]$, define an \emph{interpolating Hamiltonian} by 
$H(s) \coloneqq H_0 +s\,(H-H_0)$ and denote its thermal state by
$\gibbs_s \coloneqq \gibbs[H(s)]$. 
Then, 
\begin{equation}\label{eq:perturbation_formula}
  \begin{split}
    \Tr\bigl[A\,\gibbs_0(\beta)\bigr]
	& - \Tr\bigl[A\,\gibbs(\beta)\bigr]\\
    &= \beta\int_0^1 \rmd \tau \int_0^1 \rmd s\, \cov_{\gibbs_s(\beta)}^\tau (H-H_0,A) 
  \end{split}
\end{equation}
for any operator $A$.
\end{theorem}

The theorem holds, in particular, for all observables $A$ supported on $S$ and a perturbation $H-H_0$ only supported on $E$. 
As 
\begin{equation}\label{eq:trace_norm-expectation_values}
	\TrNorm{(\Delta\rho)^S}
	=
	\sup_{\substack{ \sNorm{A}=1\\\supp(A) = S }} \Tr[A\,\Delta \rho]
\end{equation}
for any operator $\Delta\rho$, 
the theorem indeed confirms the expectation developed before the theorem statement. 

Note that due to the factor $\beta$ on the RHS, the perturbation formula \eqref{eq:perturbation_formula} does not directly apply to ground states, as it is unclear how one can calculate the limit $\beta \to \infty$. 

The proof of Theorem~\ref{thm:perturbation_formula} is essentially a consequence of Duhamel's formula, where the key was to find the right correlation measure, the averaged generalized covariance.

\section[Applications]{Applications}\label{sec:applications}
The understanding of the correlations of thermal states and, in particular, the exponential clustering in some regimes has several applications. We review the main ones in this section.
\subsection{Thermal Lieb-Robinson bound} 
Whenever one has an explicit bound on the correlation decay in a thermal quantum system, the perturbation formula Theorem~\ref{thm:perturbation_formula} yields an explicit local stability statement. 
Now we discuss this idea explicitly for small inverse temperature $\beta < \beta^\ast$, where an exponential correlation decay is guaranteed by Theorem~\ref{thm:KGKclustering}. 
Specifically, we get back to the geometric setup sketched before Theorem~\ref{thm:perturbation_formula} of a Hamiltonian $H_0$ that is perturbed on region $E$, which is spatially separated from a subsystem of interest $S$, e.g., such as sketched in Figure~\ref{fig:covSE}. 

\begin{corollary}[Thermal Lieb-Robinson bound]
\label{cor:intensivity}
Let $S , E \subset \Vset$ be subsystems that are separated by the minimal distance, 
$\dist(S, E) \geq L_0$. 
Moreover, let $H$ and $H_0$ be a Hamiltonians with $\supp(H-H_0) \subseteq E$ and that have the same interaction graph satisfying the conditions of Theorem~\ref{thm:KGKclustering}. 
Then, for any $|\beta| < \beta^\ast$, 
\begin{equation}\label{eq:thermal_LR}
 \norm{\gibbs^S(\beta)-\gibbs_0^S(\beta)}_1  
 \leq
 \frac{ w\, |\beta|\, J } {1-\e^{-1/\xi(\beta)}} \, 
 \e^{- \dist(S,E) /\xi(\beta)} ,
\end{equation}
where $\gibbs$ and $\gibbs_0$ are the thermal states of $H$ and $H_0$, respectively, 
$w \coloneqq 4\, \min\{ |\partial S|, |\partial E| \}\,|E|/\ln(3)$, 
and $J$ bounds the interaction strengths of $H$, $H_0$, and $H-H_0$. 
\end{corollary}

This result says that, at hight temperatures, the effect on $S$ of the perturbation of $H_0$ on $E$ is exponentially suppressed in the distance between $S$ and $E$. 
In particular, this statement can be used to bound the error made when locally (on $S$) approximating a thermal state by the thermal state of a truncated Hamiltonian (see \cite[Corollary~2]{KliGogKas14} for an explicit discussion). 

This bound is reminiscent of Lieb-Robinson bounds, which can be used, e.g., to rigorously bound the error made when locally (on $S$) approximating the time evolution operator of the full Hamiltonian by a truncated one. 

The thermal Lieb-Robinson bound \eqref{eq:thermal_LR} motivates the following definition \cite{BraCra15}. 

\begin{definition}
\label{def:decay}
We say that a state $\rho$ on a spin lattice with graph $(\Vset,\Eset)$ has
\emph{$(\xi,z)$-exponentially decaying correlations}
if for any two observables $A,B$ that are normalized ($\sNorm{A}=1=\sNorm{B}$) and have disjoint support 
\begin{equation}\label{eq:def:decay}
	\cov_\rho(A,B) \leq |\Vset|^z \e^{-\dist(A,B)/\xi} \, .
\end{equation}
\end{definition}

Then Corollary~\ref{cor:intensivity} can be summarized as follows. 
Below the critical inverse temperature $\beta^\ast$ thermal states have $(\xi,z)$-exponentially decaying correlations. 

\subsection{Locality of temperature} 
%
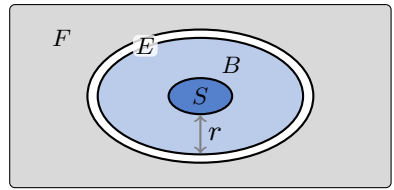
\begin{figure}
	\centering
	\begin{tikzpicture}[scale = 1.2, transform shape]
	\def \Sx{2}%
	\def \Sy{2}%
	\def \Sxsize{2}%
	\def \Sysize{2}%
	\def \Bx{8}%
	\def \By{2}%
	\tikzstyle{every node} = [inner sep = 0]
	\tikzstyle{elli} = [draw, thick, ellipse, inner sep = 0]
	\node (V) at (0,0)
		[draw, rounded corners = 2pt,
		 minimum width = 13em, minimum height = 5.5\baselineskip, 
		 fill = gray!30
		]{};

	\node (E) at (0,0) 
		[elli, minimum width = 7.7em, minimum height = 4\baselineskip, 
		 fill = white]{};    

	\node (B) at (0,0) 
		[elli, minimum width = 7em, minimum height = 3.5\baselineskip, 
		 fill = niceblue!40]{};

	\node (S) at (0,0) 
		[elli, minimum width = 2.6em, minimum height = 1.3\baselineskip, 
		 fill = niceblue, beschriftung]{$S$};

	\path (S.north) ++ (.7em,.1\baselineskip) 
			node [anchor = south west, beschriftung]{$B$}
		  (B.135) ++(3pt,-4pt) 
		  	node[anchor = south east, fill = white, opacity = .8, beschriftung, inner sep = 1pt, rounded corners = 2pt]{$E$}
		  (E.150) ++(-1.5em, 0 )
		  	node[anchor = south east, beschriftung, inner sep = 1pt, rounded corners = 2pt]{$F$};
	\draw [gray, <->, thick] (B.south) ++ (0,.7pt) -- (S.south)
		node [midway, right, text = black, inner sep = 2pt]{$r$};
	\end{tikzpicture}%
	\caption{
	Partition of the system $\Vset = B \cup E \cup F$ with $S \subset B$. 
	}
	\label{fig:buffer}
\end{figure}
%
Basic statistical mechanics teaches us that \emph{temperature} is intensive, i.e., it is a physical quantity that can be measured locally and is the same throughout the system. 
This statement holds for non-interacting systems where the global thermal state is a tensor product of the local ones. 
However, in the presence of interactions the intensiveness of temperature can break down \cite{HarMah05} and an extension of this concept is required.
For this purpose, the idea of a \emph{buffer region} was introduced \cite{Har06,FerGarAci12}. 
Here, one can take a ``ring'' $E=\{v \in \Vset: \dist(v,S) = r\}$ around $S$, which yields a partition of the total system $\Vset = B \cup E \cup F$ (disjoint union) where $B\subset S$ is the buffer region of $S$ and does not interact with $F$, see Figure~\ref{fig:buffer}. 
Next, we choose $H_0\coloneqq H_B + H_F$, where $H_B$ is the Hamiltonian only containing those interactions fully contained in $B$ and similarly for $H_F$. 
We consider $H-H_0$ as a perturbation and note that $\supp(H-H_0) = E$. 
As $\gibbs_{H_0} = \gibbs_{H_F} \gibbs_{H_B}$, 
Corollary~\ref{cor:intensivity} implies that the Hamiltonians $H$ and $H_B$ have approximately the same states on $S$, 
\begin{equation}\label{eq:local_approx}
	\gibbs_H^S(\beta) \approx \gibbs_{H_B}^S(\beta) 
	\qquad \forall \beta<\beta^\ast,
\end{equation}
up to an error exponentially small in the distance~$r$. 

As the physical application, the approximation \eqref{eq:local_approx} can be used to assign a temperature to $S$ by just knowing the state of $B$. 
This idea extends the intensiveness of temperature to non-critical interacting quantum systems. 

With respect to the critical case, in 1+1 dimensions, 
conformal symmetry completely dictates how correlation functions behave and how local expectation values of local observables of infinite systems differ from those taken for finite ones.
Hence, conformal field theory establishes that
\begin{equation}
\Tr\bigl[O\,\gibbs(\beta)\bigr] 
	 - \Tr\bigl[O\,\gibbs(\beta, H_0)\bigr]
    \simeq \frac{1}{r^y}
\end{equation}
up to higher order terms, where $y$ is the \emph{scaling dimension} of the operator $H_E$
\cite{Cardy1984, Cardy1986}. 
If $H_E$ is a standard Hamiltonian term, in the sense that the system is homogeneous, its leading scaling dimension is $y=2$.

In what follows, the same idea leads to a computational application: 
The state on $S$ can be simulated by only calculating the state on $B$. 
If one wants to simulate the global thermal state then one can resort to certain tensor network approximations discussed in the next section. 

\subsection{Classical simulatability of many-body quantum systems} 
\label{sec:class_sim}
\begin{figure}
	\centering
	\begin{tikzpicture}[scale = 1.2, transform shape,%
			anchor = center,%
	        tensorleg/.style={very thick,black!80},%
	        sa/.style = {shading = axis, shading angle = 20},%
	        site/.style = {rectangle, rounded corners = 0.5mm,%
	                       minimum height = 1.25\baselineskip, minimum width = 4.5ex,%
	                       inner sep = 2pt, , thick, draw=black!80, %
	                       top color=blue!5, bottom color=niceblue!80,sa},
	                       ]%
	  \def \nx{5}%
	  \def \slen{.25cm}%
	  \foreach \x in {0,...,\nx}{%
	  	\pgfmathsetcount{\xx}{\x+1}%
	    \node [site] (s\x) at (\x,0){$X_\the\xx$};
	    \draw [tensorleg](s\x.south) -- ++(0,-\slen) 
	    	node [anchor = north]{$j_\the\xx$};
	    \draw [tensorleg](s\x.north) -- ++(0,\slen)
	    	node [anchor = south]{$i_\the\xx$};
	    }%
	\pgfmathsetcount{\xx}{\nx}%
	  \foreach \x in {1,...,\xx}{%
	    \pgfmathsetcount{\xm}{\x-1};
	    \draw [tensorleg] (s\x.west) -- (s\the\xm.east) node[midway] (b\x){};
	  }%
	\end{tikzpicture}%
	\caption{
	Graphical MPO representation of an operator $\rho$ on a six site spin-chain. 
	Each site $v\in \Vset=[6]$ is associated with a tensor 
	$X_v \in \CC^{D(v-1)} \otimes \L(\mc H_v) \otimes \CC^{D(v)}$ 
	with \emph{bond dimensions} $D(v)$, where $D(0) = D(6) = 1$. 
	The components
	$\rho_{(i_1, i_2, \dots, i_6),(j_1,j_2, \dots, j_6)}$
	are given by the displayed tensor network, which is given by contracting the $X(v)$ over each pair of bond indices in $\{1,2, \dots, D(v)\}$. 
	}
	\label{fig:MPO}
\end{figure}
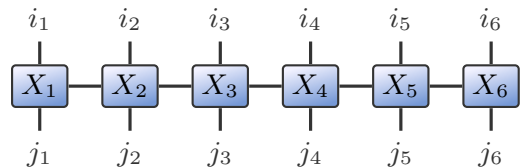
%
A naive approach for the classical simulation of many-body quantum systems
requires an amount of resources (memory and time) that scales exponentially in the
system size. 
However, in many cases, smart classical methods allow for the simulation of quantum systems by only employing polynomial resources.
A powerful family of these methods for interacting quantum systems relies on so-called \emph{tensor networks}, see \cite{Sch11} for an introduction. 
A key to the success of these methods is to find computation friendly parametrizations of quantum many-body states that have limited correlations. 
The most prominent class of states are \emph{matrix product states} (MPS), which usually represent state vectors in the Hilbert space of
$n$ spins. 
The operator analogue of MPS are \emph{matrix product operators} (MPOs), see Figure~\ref{fig:MPO} for a rough definition. 
Practically, they can be used to (approximately) represent density matrices \cite{ZwoVid04,VerGarCir04}. 

Hastings has used a certain cluster expansion method to show that thermal spin lattice states can also be approximated by MPOs \cite[Section~V]{Has06}, see also \cite[Section~III.C]{KliGogKas14}. 
After this cluster expansion method has been formalized \cite{KliGogKas14} the MPO approximation was improved and made efficient by Moln\'ar et al.~\cite{MolSchVer14}. 

For the cluster expansion one starts by expanding the exponential series of $\e^{-\beta H}$ and the Hamiltonian sum \eqref{eq:loc_H} as
\begin{equation}\label{eq:series_expansion}
	\e^{-\beta H}
	=
	\sum_{j=0}^\infty \sum_{w \in \Eset^j} \frac{(-\beta)^j}{j!} h(w) , 
\end{equation}
where $h(w) \coloneqq h_{w_1} h_{w_2} \dots h_{w_j}$. 
It can be made rigorous that correlations of range $L$ in the thermal state are due to those terms $h(w)$ where $w\in \Eset^j$ contains long connected regions in the graph $(\Vset, \Eset)$ of size at least $L$ \cite{Has06,KliGogKas14}. 
The idea underlying the MPO approximation of the thermal state is to drop those terms in the series \eqref{eq:series_expansion}. 
Then one can show that the resulting operator approximates the thermal state for small inverse temperatures $\beta < 2 \beta^\ast$ with $\beta^\ast$ from Eq.~\eqref{eq:crit_beta_def} \footnote{%
	The discrepancy of the factor of $2$ in the inverse temperature bound is due to considering two copies of the system in the proof of Theorem~\ref{thm:KGKclustering}. } 
and is an MPO, where the size of the single tensors is determined by $L$ \cite[Theorem~3]{KliGogKas14}. 
By an improved smart choice of selecting the terms in the series \eqref{eq:series_expansion} one can algorithmically obtain an improved MPO approximation to thermal states where the size of the MPO tensors scales polynomially in the system size \cite{MolSchVer14}. 

The efficient MPO approximation of thermal states \cite{MolSchVer14} naturally has implications on the simulatability of thermal states. 
Tensor network methods work best for spin-chains, i.e., in one spatial dimension. 
Here, the MPO approximation allows for efficient classical simulation including, e.g., the efficient approximation of MPO observables. 

As a last remark we comment on the positivity of MPOs. 
In general, one cannot practically check whether or not an MPO is positive-semidefinite and, hence, can be a density operator \cite{KliGroEis14}. 
However, for the thermal state MPO approximations one can easily circumvent this issue by first approximating $\e^{-\beta H/2}$ and then squaring the MPO at the cost of squaring the bond dimension \cite{Has06}.

\subsection{Phase transitions}
Expectation values and thermodynamic potentials of thermal states of finite systems are differentiable with respect to the Hamiltonian couplings and do not have non-analyticities. 
In contrast, in nature, we observe abrupt changes and discontinuities
of the thermodynamic potentials and their derivatives. 
The solution to this apparent contradiction relies in the fact that these discontinuities only appear in the limit of infinitely large system sizes, usually denoted as the thermodynamic limit.
This is actually one of the reasons why it is so difficult to prove the existence of phase transitions.

Hence, showing some behaviour for the scaling of the covariance for any system size can help to discard the existence of phase transitions.
This is the case of the clustering of correlations proven in
Theorem~\ref{thm:KGKclustering}. 
It discards the possibility of a phase transition above the critical temperature, i.e.\ where $\beta<\beta^*$ \cite{KliGogKas14}.

In this context, it is interesting to mention the existence of spontaneous magnetization at sufficiently low temperatures
$\beta > \beta_{\textrm{FSS}}$ in a relevant family of spin systems in 3 or more spatial dimensions \cite{DysLieSim76}. 
This family of models includes Hamiltonians as the XX model and the Heisenberg antiferromagnetic model with spin 1, 3/2,$\ldots$
The threshold temperature $\beta_{\textrm{FSS}}$ has a non-trivial expression given in Ref.~\cite{DysLieSim76}.
Note that the above statement simultaneously discards 
the existence of phase transitions at the very low temperature regime, $\beta < \beta_{\textrm{FSS}}$, for this type of models, 
and ensures the existence of a critical point below the disordered phase such that its Curie inverse temperature fulfills $\beta_{\textrm{FSS}}> \beta_c > \beta^*$.

\section[Energy distribution]{Energy distribution of thermal states}\label{sec:energy_distribution}
Another important property of thermal states is their energy distribution.
The mean energy of a thermal state is given by
\begin{equation}
U(T)\coloneqq \Tr\left(H \gibbs(T^{-1}) \right)\, .
\end{equation}
For or short ranged Hamiltonians this quantity scales as the volume $N = |\Vset|$. 
It is then common to use the energy density, which is defined
as the energy per site, 
\begin{equation}\label{eq:def:u}
	u(T)\coloneqq U(T)/N. 
\end{equation}

A first question is what the qualitative relation between energy and temperature is.
The \emph{heat capacity} and \emph{specific heat capacity} of the system at temperature $T$ are the quantities that tell us how much energy is necessary to increase the temperature of the system by one unit and are defined as
\begin{align}
C(T) &\coloneqq \frac{\rmd U(T')}{\rmd T'} \bigg|_{T' = T}\, , \quad&
c(T) &\coloneqq \frac{\rmd u(T')}{\rmd T'} \bigg|_{T' = T}\, , \label{eq:def:c}
\end{align}
respectively. 
A straightforward calculation leads to the identity
\begin{equation}\label{eq:HeatCapacity-EnergyFluc}
C(T) =\frac{\Delta E(T)^2}{T^2} \geqslant 0 \, ,
\end{equation}
where $\Delta E^2(T)=  \Tr[H^2 \gibbs(\beta)] - (\Tr[H\gibbs (\beta)])^2$ is the variance of the energy. 
Two salient comments regarding Eq.~\eqref{eq:HeatCapacity-EnergyFluc} are in order. 
First, the positivity of the heat capacity shows that energy is a monotonically increasing function of temperature for any system, matching with our daily life intuition.
This is in fact the reason why the energy-entropy diagrams presented in 
the book chapter \cite{NatWinLew18} 
are convex, which is a crucial property to have a non-trivial theory of thermodynamics.

Second, we can see in Eq.~\eqref{eq:HeatCapacity-EnergyFluc} that
the energy fluctuations of a thermal state
do not scale with the energy of the system but with its square root. 
In other words, the energy distribution of thermal states becomes
relatively thinner and thinner as the system size increases, and this happens irrespective of the Hamiltonian of the system.
Note that this property is necessary in order for an equivalence of
ensembles to be possible. This issue is discussed in Section~\ref{sec:EoE}.

It is interesting to point out that systems that are out of equilibrium and have a wide energy distribution cannot thermalize, that is, cannot equilibrate towards a Gibbs state. 
This is a consequence of the fact that the energy distribution of a system does not vary during the evolution and thermal states have relatively small energy fluctuations. 

If we now focus on the relevant family of states that have a finite correlation length,
it is even possible to show that the energy distribution
converges to a Gaussian with increasing system size \cite{KeaLinWel15,BraCra15}. 

Below, we state the latest result \cite{BraCra15} applied to $2$-local Hamiltonians, i.e., to Hamiltonians on cubic lattices. 
A \emph{cubic} interaction graph has a vertex set $\Vset = [n]^D$ and two sites $v,w\in [n]^D$ are connected if $\sum_{i=1}^D |v_i-w_j| = 1$; see Figure~\ref{fig:covSE} for $D=2$. 

\begin{theorem}[Berry-Esseen bound {\cite[Lemma~8]{BraCra15}}]
\label{thm:BerryEsseen}
\hfill\\
Let $H$ be a Hamiltonian \eqref{eq:loc_H} with cubic interaction graph $[n]^D$ with $N \coloneqq n^D$ sites and $\rho$ a state with $(\xi,z)$-exponentially decaying correlations \eqref{eq:def:decay}.  
Moreover, let
\begin{align}
\label{eq:defF}
F(x)&=\sum_{k:\,   E_k \le x}\langle k|\rho |k\rangle\, , \\
\mu &=\Tr(\rho H)\, ,\\
\sigma^2 &=\Tr\left( \rho (H-\mu)^2\right)
\end{align}
and define the Gaussian cumulative distribution function with mean $\mu$ and variance $\sigma^2$ by 
\begin{equation}
G(x)=\frac{1}{\sqrt{2\pi\sigma^2}}\int_{-\infty}^x\rmd y\,\medskip^{-\frac{(y-\mu)^2}{2\sigma^2}}\, .
\end{equation}
 Then
\begin{equation}  \label{eqBEmainthm}
\sup_x|F(x)-G(x)|  \le C \frac{\ln^{2d}(N)}{\sqrt{N}},
\end{equation}
where $C$ is a constant that depends only on $\xi$, $x$, and the local Hilbert-space dimension.
\end{theorem}

Hence, the distance between the cumulative distributions $F$ and
$G$ can be made arbitrarily small in $L_\infty$-norm for sufficiently large system sizes $N$. 
Note that the theorem all states with clustering of correlations, irrespective of being thermal. 

The Gaussian distribution has an exponential decay outside an energy windows of a width scaling as $\sigma \sim \sqrt{N}$. 
Hence, the theorem also suggests such a decay for states with $(\xi,z)$-exponentially decaying correlations. 
However, the error term in Eq.~\eqref{eqBEmainthm} scales like $\tilde\landauO(1/\sqrt{N})$ and, hence, such a decay is not implied. 
However, a (sub)exponential decay can be guaranteed using a version of a non-commutative Chernoff-Hoeffding inequality \cite{AraKuwLan14,Ans16,Kuw16}.

Let us now insert in Eq.~\eqref{eq:defF} the maximally mixed state $\rho =  \1 /d$, with $d$ being the dimension of the Hilbert space.
Then, $F(x)$ becomes
\begin{equation}
F(x)=\frac{1}{d}\int_{-\infty}^x \sum_k \delta(E-E_k) \, \rmd E
\end{equation}
which is nothing but the cumulative distribution of the density of states.
As the maximally mixed state is a product state, 
a corollary of Theorem~\ref{thm:BerryEsseen} is that short-ranged Hamiltonians have a density of states that tends to a Gaussian in the
thermodynamic limit.

\section{Equivalence of ensembles}\label{sec:EoE}
%
\begin{figure}
	\centering
	\includegraphics{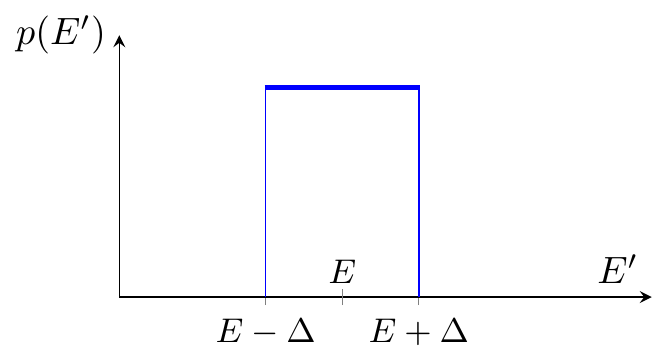}
	\caption{
		The microcanonical energy distribution. 
		The microcanonical $\micro_{E,\Delta}$ state is the corresponding maximally mixed state on energy subspace given by energies $E'$ with $|E-E'|\leq \Delta$, i.e., given by the density matrix diagonal in the energy eigenbasis and with eigenvalues given by the microcanonical energy distribution. 
	}
	\label{fig:microcanonical}
\end{figure}
%
The question of equivalence of ensembles (EoE) has a long tradition in statistical physics. 
Gibbs stated that, in the thermodynamic limit, an observable has the same average value in both the canonical and the microcanonical ensemble \cite[Chapter 10]{Gibbs1902}; 
see also Ref.~\cite{Tou15} for a recent rigorous treatment of classical systems. 
The argument given by Gibbs is that the energy fluctuations of the canonical ensemble (relatively) vanish in the limit of infinitely large systems.

In classical systems, the microcanonical ensemble describes systems at equilibrium with a well defined energy $E$. It assigns the same probability to all the configurations of the system (points in phase space) compatible with that energy. 

In quantum theory, however, for systems with a discrete energy spectrum, the classical definition of microcanonical ensemble breaks down, since there often is, if any, only a single physical configuration for a given energy.
Hence, in quantum systems the notion of microcanonical states needs to be slightly changed and the system is not considered to have a well defined energy but a uniform distribution within a narrow energy window of width $2\Delta$.  
More specifically, the microcanonical ensemble or \emph{microcanonical state} $\micro_{E,\Delta}$ of a quantum systems is defined as the completely mixed state within the energy subspace
spanned by all the eigenvectors with energy $|E_k-E|\leq \Delta$ (see Fig.~\ref{fig:microcanonical}).

Let us mention that the microcanonical ensemble is equivalent to the principle of equal a priori probabilities, which states that, once at equilibrium, the system is equally likely to be found in any of its accessible states. 
This assumption lies at the heart of statistical mechanics since it allows for the computation of equilibrium expectation values by performing averages on the phase space. However, there is
no reason in the laws of mechanics (and quantum mechanics) to suggest that the system explores its set of  accessible states uniformly. Therefore, the equal a priori probabilities principle has to be put in by hand.

One of the main insights from the field of quantum information theory to statistical mechanics is the substitution of the equal-a-priori-probabilities-postulate by the use of typicality arguments \cite{Popescu2006,GolLebTum06}. 
This issue is discussed in detail in \cite[Chapter~17]{BinCorGog19}.

An \emph{equivalence of ensembles} statement says that a certain state is locally indistinguishable from a thermal state with the right temperature. 
Such statements are often shown for the microcanonical state $\micro_{E,\Delta}$, 
see Refs.~\cite{Lim72,MueAdlMas13,BraCra15} investigations of quantum lattice systems.

\begin{theorem}[EoE {\cite[Theorem~1, special case]{BraCra15}}]
\label{thm:EoE}
Let $\gibbs(1/T)$ be the thermal state of 
a Hamiltonian~\eqref{eq:loc_H} with cubic interaction graph $[n]^D$
and with energy density $u(T)$ \eqref{eq:def:u} and specific heat capacity $c(T)$ \eqref{eq:def:c}
with 
$(\xi,z)$-exponentially decaying correlations \eqref{eq:def:decay}.

Let $\micro_{e N,\delta \sqrt{N}}$ be the microcanonical state with mean energy density $e$ specified by
\begin{equation}
	|e-u(T)| \leq \frac{\sqrt{c(T) T^2}}{\sqrt{N}}
\end{equation}
and energy spread $\delta\sqrt{N}$ specified by
\begin{equation}
	C_1\,  \frac{\ln(N)^{2D}}{\sqrt{N}}
	\leq 
	\frac{\delta}{\sqrt{c(T)T^2}}
	\leq 
	1 \, ,
\end{equation}
where $N \coloneqq n^D$ denotes the volume and $C_1$ a constant. 
Then, for a linear system size scaling as 
$n \in \tilde \Omega(l^{D+1}/\epsilon^2)$, 
\begin{equation}\label{eq:tau_approx_gibbs}
	\TrNorm{\tau_{e,\delta}^{S_l} - \gibbs(1/T)^{S_l}} \leq \epsilon  \, 
\end{equation}
for any cube $S_l$ of edge length $l$. 

All implicit constants and $C_1$ depend on the temperature $T$, the correlation decay $(\xi,z)$, and the spatial and Hilbert space dimension $D$ and $d$ (explicitly given in Ref.~\cite{BraCra15}). 
\end{theorem}

This theorem tells us that for an energy window that needs to have a width scaling roughly as 
$\Delta = \delta \sqrt N\sim\sqrt{N}$ and matching median energy 
$E =eN \approx U(T)$ 
the microcanonical state is locally indistinguishable from the thermal state whenever the total system large enough. 

The original statement \cite[Theorem~1]{BraCra15} is more general (but also more technical): 
(i) it also holds for $k$-local Hamiltonians ($k$-body interactions) and 
(ii) for systems that are not translation-invariant. 
In the case of (ii) the trace distance \eqref{eq:tau_approx_gibbs} needs to be replace with a trace distance averaged over all translates of the cube $S_l=[l]^D$. 
This means that the EoE statement holds in most regions of the non-translation invariant lattice. 

The proof \cite{BraCra15} of Theorem~\ref{thm:EoE} is based on an information theoretic result for quantum many-body systems:
 If a state $\rho$ has $(\xi,z)$-exponentially decaying correlations and $\tau$ is another state such that the relative entropy $S(\tau\Vert\rho)$ has a certain sub-volume scaling then $\tau^{S_l} \approx \rho^{S_l}$ on average over all cubes of edge length $l$. 
 The approximation error depends on the explicit scaling of $S(\tau\Vert\rho)$. 
Then a bound on $S(\micro_{E,\Delta} \Vert \gibbs(\beta))$ is proven. 
For this proof the Berry-Esseen bound Theorem~\ref{thm:BerryEsseen} is derived.

\bigskip

\section[Outlook]{Conclusions and outlook}
%
In this chapter we have reviewed several properties 
of thermal states of spin lattice Hamiltonians.
We have seen that, in a regime of high temperatures and for gapped Hamiltonians at zero temperature, equilibrium states exhibit
an exponential cluster-
This fast decay of the correlations can be exploited in several ways.
It allows for locally assigning temperature to a subsystem
even in the presence of interactions, 
as well as for approximating expectation values of local operators in infinite systems by means of finite ones.
Furthermore,  
clustering of correlations in the high temperature regime 
discards the existence of phase transitions within that regime
and implies that thermal states 
can be efficiently represented by means of matrix product operators.
We have finally seen that states with a finite correlation length
have a Gaussian energy distribution and, 
in the case that they are thermal,
cannot be distinguished by local observables 
from the microcanonical ensemble, fleshing out in such a way 
the equivalence of ensembles.

Let us conclude the chapter by mentioning the problems 
that we consider relevant in the field and still remain open. 
One of the issues where progress seems to be very difficult is
the extension of the well established theorems that logically connect area-law for the entanglement entropy, gapped Hamiltonian, and exponential clustering of correlations to a number of spatial dimensions higher than one.
We also noticed a lack of results in the direction of understanding 
how much correlated thermal states of local Hamiltonians can be. 
More explicitly, how far concrete models are from saturating the area law for the mutual information.
While there has been an important effort to contrive 
local Hamiltonians with highly entangled ground states, 
little is known for thermal states.

Completely new physics is expected to appear if one goes beyond the assumption of short ranged interactions. 
In this respect, 
a class of relevant models that both appears frequently in nature and can be easily engineered, e.g.\ in ion traps, are systems with long range interactions.
For those, most of the questions discussed above are unexplored.
As it already happened for the Lieb-Robinson bounds \cite{Gong2014,Hauke2013,Matsuta2017}, 
some of the above properties, e.g.\ some type of clustering of correlations, could, at least in spirit, be reproduced. 
Indeed, for fermionic spin lattices an algebraic correlation decay can be shown for power-law interactions at non-zero temperature \cite{HerGogCir17}. 

An interesting feature of systems with long range interactions is that they do not have a finite correlation length and thereby the results on equivalence of ensembles presented above cannot be applied. 
This opens fundamental questions in the field of equilibration of closed quantum systems, since it is not obvious that the canonical ensemble will be an appropriate description of the equilibrium state \cite{Casetti2007}. 
The challenge of the equivalence of ensembles for states with 
a power law decay of the correlations yields another relevant question in the field. 

Other than extending EoE statements to long range interactions it is also important to extend them to larger classes of states. 
Often, states in quantum experiments arise from quenched dynamics and is a largely open and challenging problem to characterize the thermalization behaviour. 
Part of the problem is the equivalence of the equilibrium state (given by an infinite time average \cite{Popescu2006}) and the thermal state.
For non-critical systems such a statement can indeed be proven \cite{FarBraCra17}.

Thermal states have been introduced at the beginning of this book chapter as
states that maximize the von Neumann entropy given the energy as conserved quantity. 
However, in many relevant situations, in particular for integrable models evolving completely isolated from their environments (closed system dynamics), it is well 
known that the thermal state is not a good description of the equilibrium state. 
These type of systems require to consider additional conserved quantities
and their equilibrium state is given by the so called Generalized Gibbs Ensemble (GGE),
which is the state maximizing the entropy given a set of conserved quantities. 
In the recent years, several studies on the laws of thermodynamics
when GGE states are taken as heat baths have been realized (see 
the book chapter \cite{MinGurFai18}).
However, their intrinsic properties are in general completely unknown: scaling of the correlations, the shape of the energy distribution, or the locality of the Lagrange multipliers (e.g.\ temperature and chemical potential). 
Characterizing these properties constitutes an extensive research endeavor.

\section{Acknowledgments}
The work of MK was funded by the National Science Centre, Poland (Polonez 2015/19/P/ST2/03001) within the European Union's Horizon 2020 research and innovation programme under the Marie Sk{\l}odowska-Curie grant agreement No 665778.
AR acknowledges Generalitat de Catalunya (AGAUR Grant No. 2017 SGR 1341, SGR 875 and CERCA/Program), the Spanish Ministry MINECO (National Plan 15 Grant:FISICATEAMO No. FIS2016-79508-P, SEVERO OCHOA No. SEV-2015-0522), Fundaci\'o Cellex, ERC AdG OSYRIS, EU FETPRO QUIC, and the CELLEX-ICFO-MPQ fellowship.


%

\end{document}